\documentclass[a4paper]{spie}  

\usepackage{amsmath,amsfonts,amssymb}
\usepackage{graphicx}
\usepackage[colorlinks=true, allcolors=blue]{hyperref}
\usepackage{stmaryrd}
\usepackage{subcaption}
\usepackage{multirow}

\newfont{\cyr}{wncyr10 scaled 1200}%
\newcommand{\shuffle}{{\mbox{\cyr sh}}}
\title{Optimal exposure time for satellite imaging? A trade-off between residual tip-tilt and registration accuracy}

\author[a]{Florian Cheyssial}
\author[a]{Laurent M. Mugnier}
\author[a]{Cyril Petit}
\affil[a]{DOTA, ONERA, Université Paris Saclay, BP 72, 92322 Châtillon cedex, France}

\authorinfo{Send correspondence to F.C. Email = firstname.lastname@onera.fr}

\usepackage{fancyhdr}
\usepackage{lastpage}
\renewcommand{\headheight}{25pt} 
\renewcommand{\headsep}{40pt} 

\begin{document} 
\maketitle

\begin{abstract}
Adaptive-Optics assisted astronomical imaging and satellite imaging from the ground share many commonalities but have also specificities. In particular satellites are non-stationary objects in terms of both shape and flux, encouraging relatively short exposure times. Besides, very short exposures freeze the turbulence and preserve high spatial frequency information in speckles, but at the cost of a low image SNR, which complicates the post-processing. Conversely, long exposures lead to speckle-less images where some high frequency information is lost. In this communication, we propose a method for determining the exposure time that optimizes the image quality after post-processing. In particular, we focus on the reduction of the jitter effects --~induced by atmospheric tip-tilt residues, vibrations and object non stationarity~-- which may dramatically reduce the image quality when imaging satellites. Our model takes into account the object type and flux, the residual aberrations power spectral density and the noise level. 
\end{abstract}


\section{INTRODUCTION}
\label{sec:intro}  

Space domain awareness (SDA) has become a burning issue due to the increasing number of satellites in orbit and, more broadly, due to the growth of space activities. SDA, amongst other things, encompass the detection, characterization and monitoring of near-Earth space objects (including satellites, debris and asteroids). This can be achieved through direct imaging from the ground, using optical telescopes. In particular, direct imaging is well suited to characterization and monitoring, for which measurements with high angular resolution are required. Ground-based observation is severely limited by the presence of atmospheric turbulence which distorts the wavefronts incident to the telescope, and hence reduces the angular resolution of the acquired images. Adaptive-optics (AO)-systems can be used to mitigate the atmospheric turbulence effects, which drastically improves the resolution, but the correction is only partial and image restoration (\textit{i.e.}, deconvolution) is often required to further improve it.

We consider an AO-assisted imaging system that acquires a series of images of either a satellite, debris or asteroid. These images are then jointly processed to form a restored image, the quality of which we aim to maximize. For such optical-digital systems imaging through turbulence, one of the key parameters affecting the restored image quality is the exposure time. On the one hand, increasing the exposure time increases the intensity of the focal plane images, and therefore improves the images signal-to-noise ratio (SNR), which is beneficial to the image restoration process. On the other hand, it also reduces the high spatial frequency content of the focal plane images because of the integration of the speckles in the point spread function (PSF). Conversely, as the exposure time decreases, turbulence is ``frozen'' and the focal plane images preserve more high spatial frequency information, but the image SNR decreases, which complicates the image restoration.

When imaging satellites, debris or asteroids, an important factor in the broadening of the PSF as the exposure time increases is the integration of the PSF jitter, induced mostly by atmospheric tip-tilt residues and vibrations. In this case, a solution to minimize the PSF width while maintaining a high number of photons for the deconvolution is to: acquire several images with very short exposure time, and to register the image stack (\textit{i.e.}, performing a shift \& add process), to compensate for the average shift of each image. Deconvolution can then be performed on the registered image to correct for the remaining aberrations. The quality of the registered image depends on: the quantity of jitter integrated during a single frame (which increases as the exposure time increases), and on the registration errors (which increases as the exposure time decreases). 

In this paper, we present a method to optimize the exposure time of an AO-assisted imaging system for satellite imaging, with respect to the combined effects of integrated jitter and registration errors. It takes into account the turbulence conditions, the object flux and shape, and the noise level. The paper is organized as follows. We first discuss the trade-off between jitter integration and image registration errors in Sect.~\ref{sec:tradeoff}. In Sect.~\ref{sec:model}, we derive a model of the registered image that highlights the impact of integrated jitter and registration errors on its resolution. As an illustration, the model is used to optimize the exposure time in the case of LEO satellite observation in Sect.~\ref{sec:results}. We discuss the method and conclude on this work in Sect.~\ref{sec:ccl}.

\section{Trade-off between integrated jitter \& registration error}\label{sec:tradeoff}

We consider an AO-assisted telescope that disposes of a period $T$ to observe a satellite, such that the satellite remains stationary during this time. To make the most of the available photons during this period without being affected by the integration of the image jitter, the period $T$ is divided into a series of short exposure images --~such that the jitter integrated in each frame is minimized~-- and the images are registered. The registration consists in estimating the average shift of each image, before aligning and summing them to obtain an image with the same intensity as if it were integrated over $T$. The registered image can then be deconvolved to correct for the remaining aberrations.

\begin{figure}[h]
   \centering
   \begin{subfigure}[t]{0.65\textwidth}
      \centering
        \includegraphics[width=\textwidth]{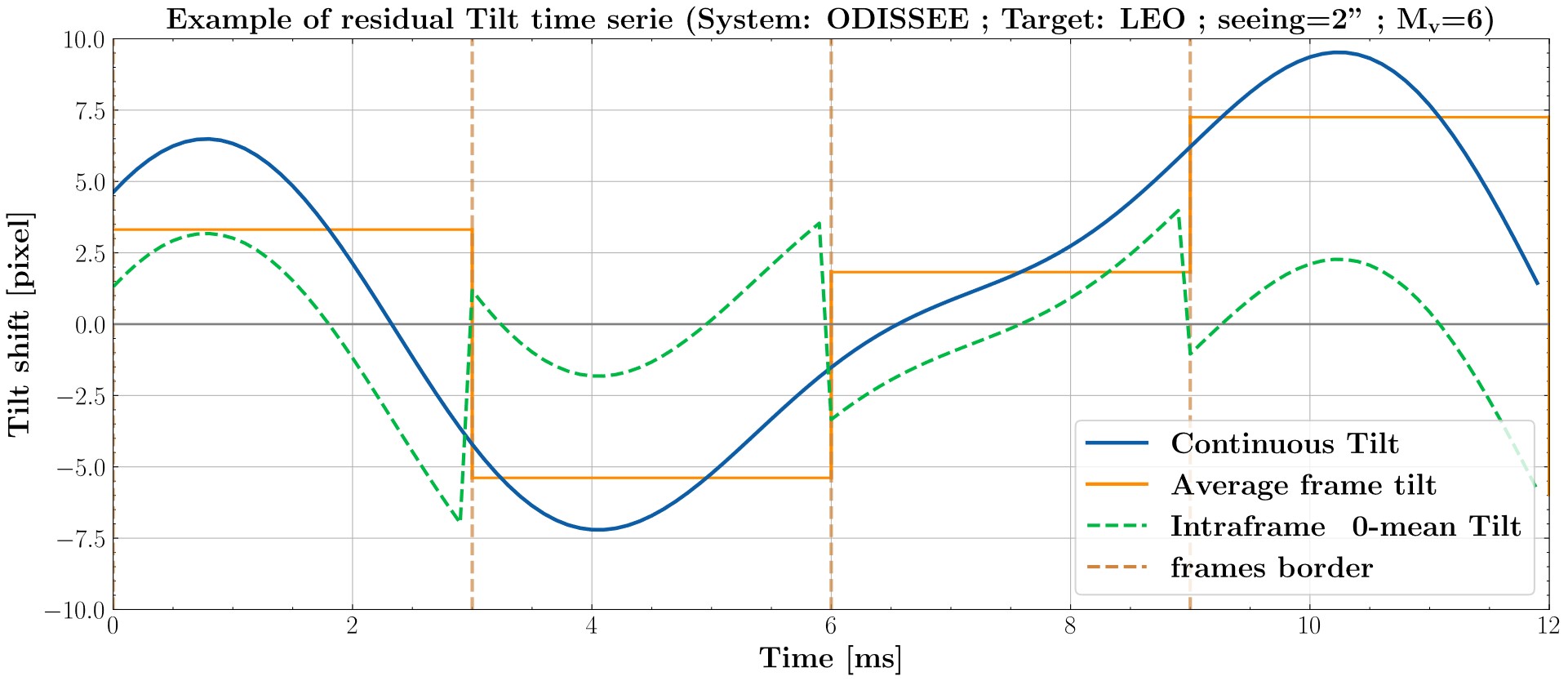}
        \caption{\textbf{Blue solid curve}: Simulation of an atmospheric residual tilt realization. \textbf{Green dashed line}: Tilt after registration. This simulation was obtained under the conditions of a LEO satellite observation using an ODISSEE-like system, as described in Sect.~\ref{sec:results}.}
        \label{fig:tilt_real}
   \end{subfigure}
   ~
   \begin{subfigure}[t]{0.30\textwidth}
      \centering
        \includegraphics[width=\textwidth]{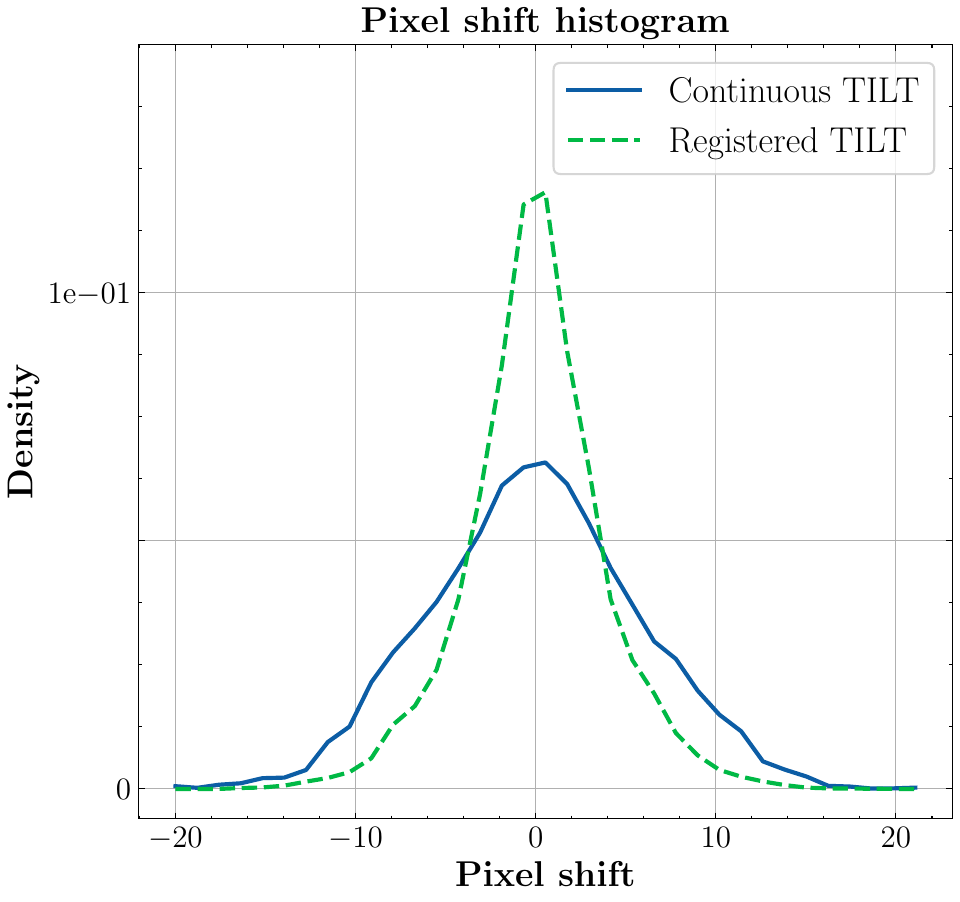}
        \caption{\textbf{Blue solid curve}: normalized histogram of the tilt realization. \textbf{Green dashed curve}: normalized histogram of the registered tilt.}
        \label{fig:tilt_histo}
   \end{subfigure}
   \caption{}
   \label{bbb}
\end{figure}

To illustrate this principle, we plot on Fig.~\ref{fig:tilt_real} a simulated realization of atmospheric residual tilt as a function of time (see the blue solid curve). The dashed orange vertical lines delimit the short-exposure frames borders, the average tilt in each frame is given by the orange step function and the green dashed curve corresponds to the registered tilt. The registered tilt is simply the residual tilt (blue curve) minus its average value in each frame (orange curve). If a single long-exposure frame is acquired, the integration of the tilt results in a blur equivalent to a convolution with the tilt histogram (see the solid blue curve in Fig.~\ref{fig:tilt_histo}). In the registered image however, the blur induced by the integration of the jitter is given by the histogram of the registered tilt, which is narrower than the histogram of the non-registered tilt (compare the two curves in Fig.~\ref{fig:tilt_histo}). Dividing the observation into short-exposures and registering the images filters the integrated jitter.

Provided the registration is perfect, reducing the exposure time reduces the amount of jitter integrated within each frame, making the registered tilt histogram narrower and the registered image sharper. However, the precision of the average shift estimation for each frame decreases as the frame SNR decreases, and therefore as the exposure time decreases. Misestimating the average shift of the images introduces a blur to the registered image that is similar to the blur caused by jitter integration. Consequently, there should exist an optimal exposure time for which the combined effects of the integrated jitter and of the registration errors are minimized. 




\section{Registered image model}\label{sec:model}

In this section, we develop a model to describe the registered image as a function of the exposure time.

\subsection{Image formation model}\label{ssec:image}

To describe the registered image, we first need to describe the short-exposure images acquired at the focal plan of the AO-assisted telescope. We denote $\left\{ \mathbf{i_k} \right\}_{k\in\llbracket 1,N_f\rrbracket}$ the set of $N_f$ frames acquired during the total period $T$, each with an exposure time $\tau$. Each one of these images can be expressed as follows:
\begin{equation}
   \forall k\in\llbracket 1,N_f\rrbracket: \mathbf{i_k} = N_{ph}\left[ \langle h_{opt}\rangle_{\tau,k} \ast h_{det} \ast o\right]_\shuffle + \mathbf{n_k},
   \label{eq:image_model}
\end{equation}
where $N_{ph}$ is the number of photon in each image ($N_{ph}\propto\tau$), $o$ is the observed object, $h_{det}$ is the detector's response, $h_{opt}$ is the instantaneous optical PSF and $\mathbf{n_k}$ is the noise. Italic letters are used to represent continuous functions and scalars, while bold letters are used to represent discrete value vectors. The symbols $\ast$ and $\shuffle$ respectively stand for the convolution and discretization operators. The notation $\langle h_{opt}\rangle_{\tau,k}$ denotes the integration of $h_{opt}$ between the times $k\tau$ and $(k+1)\tau$. Both the object $o$, the detector's PSF $h_{det}$ and the integrated optical PSF $\langle h_{opt}\rangle_{\tau,k}$ are normalized such that the image intensity is only given by $N_{ph}$. The noise term $\mathbf{n_k}$ is assumed to be a mixture of Poisson and Gaussian noise, to account for both photon and readout noise.

The instantaneous optical PSF, $h_{opt}$, can be decomposed into a jitter kernel, $h_{jit}$, and a jitter free kernel, \textit{i.e.}:
\begin{equation}
   h_{opt} = h_{jit} \ast h_{ttr}.
\end{equation}
Here, $h_{jit}$ is simply the instantaneous shift of the optical PSF, which takes the form of a Dirac whose position evolves with time. This includes atmospheric residual tip-tilt, vibrations and any other source of jittering. The other term $h_{ttr}$ (for tip-tilt Removed) contains all the higher order aberrations and the diffraction (\textit{i.e.}, this is the centred instantaneous PSF speckle pattern). 

\subsection{Expression of the registered image}\label{ssec:registration}

The image stack $\left\{ \mathbf{i_k} \right\}_{k\in\llbracket 1,N_f\rrbracket}$ is then registered, \textit{i.e.}: the shift of each image is estimated, then the images are recentred and summed. The registered image, denoted $\mathbf{i_\Sigma}$, takes the following form: 
\begin{equation}
   \mathbf{i_\Sigma} = \sum^{N_f}_{k=1}\left[ \delta\left(  x+\hat{\alpha}^k_x, y+\hat{\alpha}^k_y\right)\ast \mathbf{i_k}\right]_\shuffle,
\end{equation} 
where $\delta$ is a dirac function, $x$ and $y$ are the two-dimensional space coordinates, and $\{\hat{\alpha}^k_x, \hat{\alpha}^k_y\}$ are the shifts estimated for the $k^{th}$ frame along the $x$ and $y$ axis. To estimate the shift between the images, we use a maximum-likelihood (ML) approach as described in Ref.~\citenum{gratadourSubpixelImageRegistration2005}. Assuming that the number of frames $N_f$ is large enough for the sum to converge, we approximate the registered image as follows:
\begin{equation}
    \mathbf{i_\Sigma} \approx N_{ph}^{tot}\left[ h_{reg}\ast \overline{\langle h_{jit}\rangle_\tau^c}\ast \langle h_{ttr}\rangle_\infty\ast h_{det}\ast o \right]_\shuffle + \mathbf{n_\Sigma},
    \label{eq:approxreg}
\end{equation}
%
where $N_{ph}^{tot}=N_f\times N_{ph}$ is the total number of photon in the registered image, and $\mathbf{n_\Sigma}$ is the registered noise, equal to: $\sum^{N_f}_{k=1}\left[ \delta\left(  x+\hat{\alpha}^k_x, y+\hat{\alpha}^k_y\right)\ast \mathbf{n_k}\right]_\shuffle$. The first term in the brackets, $h_{reg}$, is a blurring kernel that captures the effects of the registration errors on the registered image. It is defined as the distribution of the registration errors. The second term, $\overline{\langle h_{jit}\rangle_\tau^c}$, is the average of the jitter integrated within each frame and recentred. In other words, it corresponds to the normalized histogram of the registered jitter, as illustrated in Sect.~\ref{sec:tradeoff} for one dimension (see the green dashed curve in Fig.~\ref{fig:tilt_histo}). This kernel accounts for the impact of the jitter integration within each frame. The last kernel, $\langle h_{ttr}\rangle_\infty$, is the integration of the tip-tilt removed PSF over an infinitely long exposure time. The separation of the jitter and tip-tilt removed contributions relies on the assumption that their correlation can be neglected. 

In the simplified expression of Eq.~\eqref{eq:approxreg}, the resolution of the registered image is dictated by the equivalent PSF $h_\Sigma = h_{reg}\ast \overline{\langle h_{jit}\rangle_\tau^c}\ast \langle h_{ttr}\rangle_\infty\ast h_{det}$. Amongst these terms, only two depends on the exposure time: the registration error kernel, $h_{reg}$, that gets wider as the exposure time decreases; and the integrated jitter kernel, $\overline{\langle h_{jit}\rangle_\tau^c}$, that gets wider as the exposure time increases. In the two following subsection, we derive a model for these two kernels. 

\subsection{Expression of the integrated jitter kernel $\overline{\langle h_{jit}\rangle_\tau^c}$}\label{ssec:hjitter}

The integrated jitter kernel, $\overline{\langle h_{jit}\rangle_\tau^c}$, is defined as the histogram of the registered jitter, where the registered jitter is the image jitter to which we subtract the jitter average value for each frame (see the green dashed curve on Fig.~\ref{fig:tilt_real}). As a first approach, we assume that the registered jitter has an isotropic distribution, and we model $\overline{\langle h_{jit}\rangle_\tau^c}$ as a centred isotropic Gaussian function of width $\sqrt{\sigma_{jit}^2}$, where $\sigma_{jit}^2$ is the variance of the registered jitter. It corresponds to the quantity of jitter that is not filtered by the registration process. This variance is computed as the average of the registered jitter variance along the $x$ and $y$ axis. In the following, we present the calculation of the registered jitter variance along one arbitrary axis.

Assuming that the jitter is a stationary random process characterized by its temporal power spectral density (PSD) $S_{jit}$, the variance $\sigma_{jit}^2$ can be calculated as:
\begin{equation}
   \sigma_{jit}^2 = \int S_{jit}(\nu)\left( 1 - \mathrm{sinc}_\pi^2(\tau\nu) \right)\mathrm{d}\nu,
\end{equation}
where $\nu$ is the temporal frequency. The temporal PSD of the jitter can be decomposed as a sum of the temporal PSD of the different jitter contributions. For instance, if we consider only atmospheric residual tip-tilt and vibrations, the jitter temporal PSD is decomposed as follows:
\begin{equation}
   S_{jit} = S_{tip,tilt} + S_{vib},
\end{equation}
where $S_{tip,tilt}$ is the temporal PSD of the jitter induced by either atmospheric residual tip or tilt (depending on the axis), and $S_{vib}$ is the temporal PSD of the jitter caused by vibrations. The impact of vibrations on the image jitter depends greatly on the telescope structure and vibration sources. Therefore, $S_{vib}$ is often obtained by direct measurement or complete simulations based on finite element methods (see for example Ref.~\citenum{SedghiAnalyzingImpactVibrations2016,leon-gilVibrationManagementGroundbased2025}). Regarding $S_{tip,tilt}$, it is equal to the temporal PSD of the residual tip-tilt coefficients, with a scaling factor for the conversion into pixel units. To compute the temporal PSD of the residual tip-tilt coefficients, we use the method proposed in Ref.~\citenum{conanWavefrontTemporalSpectra1995}. First, we simulate the spatial PSD of the residual phase for each layer of the turbulence profile. From these phase PSDs, we can then compute the spatial PSD of the tip-tilt coefficients for each layer using the following relation:
\begin{equation}
   W_{tip,tilt}^l(\vec{f}\ ) = \left| \tilde{Z}_{tip,tilt}(R\vec{f}\ ) \right|^2 W_{\varphi}^l(\vec{f}\ ),
\end{equation}
where $\vec{f}$ is the 2D spatial frequency vector, $R$ is the radius of the telescope and $\tilde{Z}_{tip,tilt}$ is the Fourier transform of the tip, or tilt, Zernike mode (defined on a unit disk). The two functions $W_{\varphi}^l$ and $W_{tip,tilt}^l$ are, respectively, the spatial PSD of the $l^{th}$ layer residual phase, and the spatial PSD of the tip, or tilt, coefficient for the $l^{th}$ layer. Using Tyler's frozen flow hypothesis, the spatial PSDs of the tip and tilt coefficients for each layer can be converted to temporal PSDs via the following relation:
\begin{equation}
   S_{tip,tilt}^l(\nu) = \iint W_{tip,tilt}^l(\vec{f}\ )\delta(\vec{v}_l\cdot\vec{f}-\nu)\mathrm{d}^2\vec{f},
\end{equation}
where $\vec{v}_l$ is the wind vector of the $l^{th}$ layer, $\cdot$ denotes the scalar product and $S_{tip,tilt}^l$ is the temporal PSD of the tip, or tilt, of the $l^{th}$ layer. Finally, the total PSD $S_{tip,tilt}$ is the sum of all the layer's contributions. 

\subsection{Expression of the registration error kernel $h_{reg}$}\label{ssec:hreg}

As for the integrated jitter kernel, we assume that the registration error distribution is isotropic, and we model the registration error kernel, $h_{reg}$, by a centred isotropic Gaussian of width $\sqrt{\sigma_{reg}^2}$, where $\sigma_{reg}^2$ is the registration error variance. While the variance of the registration error can be obtained empirically using simulations, deriving an analytical expression is not straightforward. As an alternative, we propose to use the Cramer-Rao lower bound (CRLB) as a proxy, since its expression can be derived analytically. The CRLB, denoted $\sigma_{crlb}^2$, is such that $\sigma_{reg}^2\ge\sigma_{crlb}^2$, as we considered that the shifts of the focal plane images were estimated using ML estimation, which is unbiased. In Sect.~\ref{sec:results}, we discuss the impact of using the CRLB instead of computing $\sigma_{reg}^2$. 

As a first approach to derive an expression of the CRLB, we use a simplified model to describe the formation of the focal plane images. We assume that the focal plane images stack $\{\mathbf{i_k}\}_{k\in\llbracket1,N_f\rrbracket}$ consists of images that differ only in terms of shift and noise, \textit{i.e.}:
\begin{equation}
   \forall k\in\llbracket1,N_f\rrbracket,\ \mathbf{i_k} =  \left[g\ast\delta\left(  x-\alpha^k_x, y-\alpha^k_y\right)\right]_\shuffle + \mathbf{n_k},
   \label{eq:crlbimmod}
\end{equation}
where $\{\alpha^k_x,\alpha^k_y\}$ is the true shift of the $k^{th}$ image, $g$ is the noiseless reference image, and $\mathbf{n_k}$ is a mixture of Poisson and Gaussian noise, as in Eq.~\eqref{eq:image_model}. In our case, the noiseless reference image, $g$, is defined as follows:
\begin{equation}
   g = N_{ph}\left( \overline{\langle h_{jit}\rangle_\tau^c}\ast \langle h_{ttr}\rangle_\infty \ast h_{det} \ast o\right).
\end{equation}
For $\mathbf{n_k}$ being a mixture of Gaussian and Poisson noise, it is shown in Ref.~\citenum{barrettMaximumlikelihoodMethodsWavefront2007} that a good approximation for the Fisher information matrix (FIM), denoted $F$, is : 
\begin{equation}
   \forall(s,l)\in\llbracket1,N_\theta\rrbracket^2\quad :\quad F_{s,l} \approx \sum_{k=1}^{N_f}\sum_{m=1}^{N_{pix}}\frac{1}{\sigma_{ron}^2+g_k[m]}\,\frac{\partial g_k[m]}{\partial\theta_s}\,\frac{\partial g_k[m]}{\partial\theta_l}
    \label{eq:fim}
\end{equation}
where: $N_{pix}$ is the number of pixels in the images; $\sigma^2_{ron}$ is the variance of the Gaussian noise component (here it corresponds to the readout noise variance); $\mathbf{g_k}$ is the noiseless $k^{th}$ frame, defined as $\left[g\ast\delta\left(  x-\alpha^k_x, y-\alpha^k_y\right)\right]_\shuffle$; and $\{\theta_s\}_{s\in\llbracket1,N_\theta\rrbracket}$ is the set of unknown parameters of the model. Here, we consider that the set of unknown parameters only contains the shifts of the focal plane images, and therefore  that the reference $g$ is known, which is a strong assumption (in practical cases, the reference is unknown). The impact of this assumption is discussed in Sect.~\ref{sec:results}. Finally, the CRLB on the estimation of the different shifts is given by the diagonal terms of the inverse of the FIM.

\section{Minimization of the total PSF width}\label{sec:results}

Based on the model that we derived for the registered image (see Eq.~\eqref{eq:approxreg}), the combined effects of jitter integration and registration errors on the registered image are captured by the total kernel $h_{tot}=h_{reg}\ast \overline{\langle h_{jit}\rangle_\tau^c}$, which is a Gaussian kernel of width $\sigma_{tot}=(\sigma_{jit}^2+\sigma_{crlb}^2)^{1/2}$. Therefore, the optimal exposure time with respect to these combined effects can be defined as the exposure time that minimizes $\sigma_{tot}$. To calculate it, we compute $\sigma_{tot}$ as a function of the exposure time, using the formulas of Sect.~\ref{ssec:hjitter} and Sect.~\ref{ssec:hreg}, and look for the minimum.

\begin{figure}[htbp]
   \centering
   \includegraphics[width=0.75\textwidth]{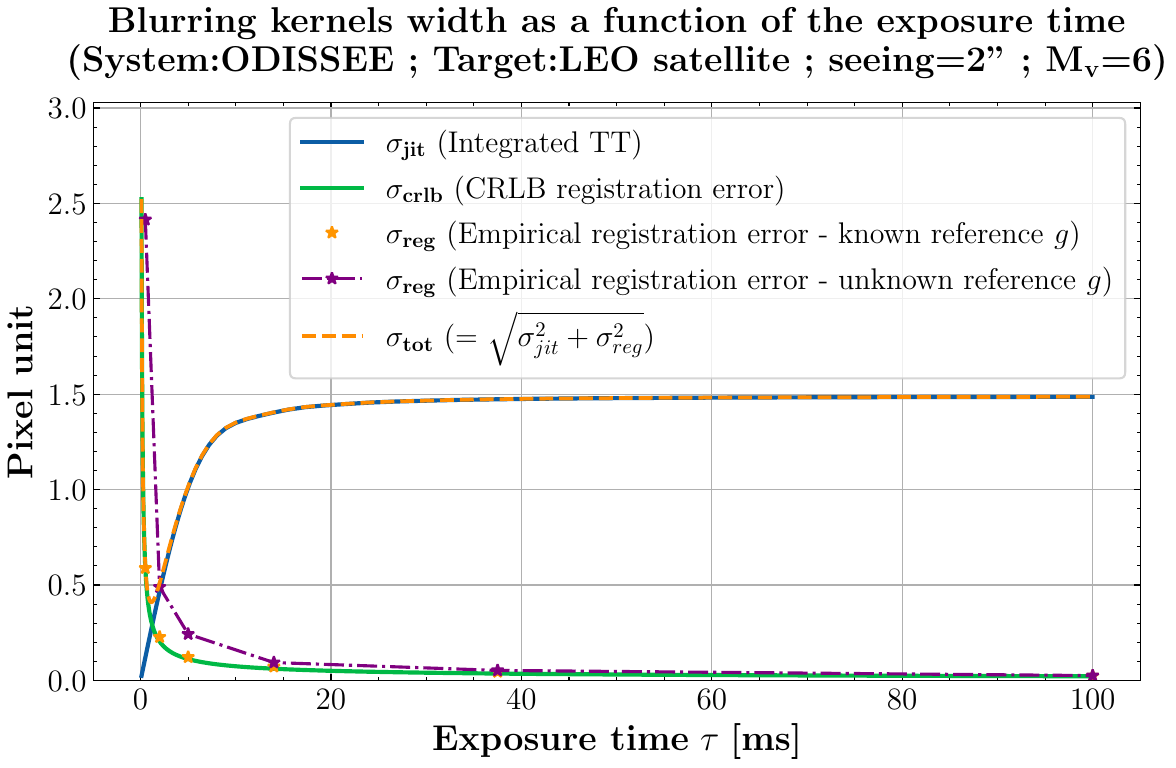}
   \caption{\centering Registration error and integrated jitter standard deviations as a function of the exposure time for the observation of a LEO satellite (seeing: $2"$, Object magnitude in V band: 6)}
   \label{fig:topt}
\end{figure}

To illustrate our method, we perform the calculation of $\sigma_{jit}$, $\sigma_{crlb}$ and $\sigma_{tot}$ as a function of the exposure time, in the case of an observation of a LEO satellite orbiting at 800\,km. The telescope and AO-system designs used for the calculation are based on the ODISSEE system design \cite{petitLEOSatelliteImaging2020}, whose key features are listed in Tab.~\ref{tab:params}. To describe the turbulence, we consider a profile with two layers, whose characteristics are given in Tab.~\ref{tab:turbulence}. The first layer accounts for turbulence at low altitudes, where it is strongest, whilst the second layer accounts for turbulence at high altitudes, which is weaker but moves quickly due to satellite tracking. For the calculation of the integrated jitter variance, we only consider the jitter induced by atmospheric residual tip-tilt (in particular, we do not consider the vibrations). The spatial PSDs of the residual tip-tilt are simulated using a Fourier-based AO-simulation tool called ``AOerror", the principle of which is explained in Ref.~\citenum{fetickIncludingPyramidOptical2023}. The results are presented in Fig.~\ref{fig:topt}, for a turbulence seeing of $2"$ and an object magnitude in the V band of 6 (which gives a pixel flux around 4700 photons per second for the satellite we considered). It can first be observed that an optimal exposure time exists, \textit{i.e.}, $\sigma_{tot}$ (orange dashed curve) has a minimum, whose value is around 1\,ms. For shorter exposures, $\sigma_{tot}$ is dominated by the registration errors (green solid curve), while for greater exposures it is dominated by the jitter integration (blue solid curve). It can be seen that, beyond $\approx10$\,ms, $\sigma_{jit}$ (and hence $\sigma_{tot}$) ceases to increase with exposure time. This is because the low temporal frequencies of the atmospheric tip-tilt are corrected by the AO-system, thus the residual tip-tilt is mostly high frequency. As a result, once the exposure time exceeds a certain threshold (here around 10\,ms), all the jitter statistic is integrated. Past this threshold, there is no interest in making registration. Compared to a long pose, jitter effects are reduced by $\approx66\%$ at the optimal exposure time.

\begin{table}[htbp]
\caption{Telescope and AO-system key parameters}
  \label{tab:params}
  \centering
   \begin{tabular}{cccc}
      \hline
      & Parameter & Value & Unit\\
      \hline\hline
      \multirow{2}{10em}{Telescope design} & Diameter & 1.5 & [m] \\
      & Central obstruction & 37.5 & [cm] \\
      \hline
      \multirow{6}{10em}{AO system design} & loop frequency & 1500 & [Hz] \\
      & Number of sub-apertures of Shack-Hartman wavefront sensor & 8$\times$8 & $\varnothing$ \\
      & Number of actuators of the deformable mirror & 9$\times$9 & $\varnothing$ \\\
      & Wavefront sensor spectral bandwidth & [500,800] & [nm]\\
      & Imaging spectral bandwidth & [675,725] & [nm]\\
      & Readout noise (imaging channel) & 1.4 & [$e^{-}$/pixel]\\
      \hline
   \end{tabular}
\end{table}

\begin{table}[htbp]
\caption{Turbulence profile parameters (wind speed including apparent wind induced by satellite tracking)}
  \label{tab:turbulence}
  \centering
   \begin{tabular}{cccc}
      \hline
      & Parameter & Value & Unit\\
      \hline\hline
      \multirow{3}{10em}{First layer} & altitude & 0 & [km] \\
      & $Cn^2$ ratio & 0.8 & $\varnothing$ \\
      & wind speed & 6 & [m.$\mathrm{s}^{-1}$] \\
      \hline
      \multirow{3}{10em}{Second layer} & altitude & 10 & [km] \\
      & $Cn^2$ ratio & 0.2 & $\varnothing$ \\
      & wind speed & 110 & [m.$\mathrm{s}^{-1}$] \\
      \hline
   \end{tabular}
\end{table}

To make a comparison with the CRLB, we calculate empirically the standard deviation of the registration error, $\sigma_{reg}$, for two cases. In the first case, the reference image $g$ is known, as it is supposed in the calculation of the CRLB, whilst in the second case, the reference is estimated jointly with the shifts. In both cases, the image stack used to compute the registration error is simulated using the formula of Eq.~\eqref{eq:crlbimmod}. When the reference is known, ML estimation reaches the CRLB, \textit{i.e.}, $\sigma_{reg}=\sigma_{crlb}$ (compare the green solid curve and the orange stars in Fig.~\ref{fig:topt}). This makes the CRLB a suitable proxy in this case. However, when the reference image needs to be estimated, the registration error increases. This is particularly evident at low exposure times, when the image SNR is poor. By extension, the optimal exposure time also increases, rising from 1\,ms to 2\,ms. Therefore, the current model slightly underestimates the optimal exposure time in real cases where the reference image must be estimated. Furthermore, if we consider a more realistic image stack in which the PSF varies between frames (see Eq.~\eqref{eq:image_model}), we can expect the registration error (and hence the optimal exposure time) to increase further.


\section{Discussion \& conclusion}\label{sec:ccl}

We proposed a method to optimize the exposure time of a system that acquires a series of images, which are then registered, with regard to the combined effects of integrated jitter and registration errors. This method is based on a simplified model of the registered image, which highlights the impact of integrated jitter and registration errors on its resolution. Various simplifications and assumptions have been made and not developed here. The model takes into account the turbulence conditions, the object flux and shape, and the noise level. For illustration, the method was applied to the case of a LEO satellite observation, considering only residual tip-tilt induced jitter. In particular, it was shown that an optimum does exist. 

It is important to note that the proposed method is merely a first step towards optimizing exposure time and there is room for a number of improvements. Firstly, the distribution of the jitter is assumed to be isotropic, which is dubious in practice, especially when observing satellites, in which case the PSF is elongated in the direction of the satellite's movement. Secondly, the CRLB tends to underestimate the variance of the registration error in real conditions where the PSF varies between the frames and the reference image is unknown. To address this issue, future works should include the reference estimation in the calculation of the CRLB. With regard to the issue of PSF variations, one possible solution would be to model these variations as noise, so that they can be taken into account in the CRLB calculation. Finally, the proposed method only optimizes the exposure time with regard to the trade-off between jitter integration and registration error. In particular, the SNR of the registered image is not taken into account in the optimization of the exposure time. A significant enhancement to the method would be to replace the registration with a multi-frame blind deconvolution algorithm, which includes the estimation of both the shift and the higher order PSF modes in each image, and to optimize the exposure time regarding the quality of the deconvolved image. Considering all the current limitations of the model, the value of the optimal exposure time obtained in Sect.~\ref{sec:results} should be viewed with a degree of detachment.

That being said, the current method can still be used to understand the impact of the AO system design, turbulence conditions and target type on the optimal exposure time, or to determine a lower bound for it.

\acknowledgments 
 
We acknowledge support from project PEPR Origins, reference ANR-22-EXOR-0016,
supported by the France 2030 plan managed by Agence Nationale de la Recherche.

\bibliography{report} 

\begin{thebibliography}{1}

\bibitem{gratadourSubpixelImageRegistration2005}
Gratadour, D., Mugnier, L.~M., and Rouan, D., ``Sub-pixel image registration
  with a maximum likelihood estimator - {{Application}} to the first adaptive
  optics observations of {{Arp}} 220 in the {{L}}{$\prime$} band,'' {\em
  Astronomy \& Astrophysics}~{\bf 443},  357--365 (Nov. 2005).

\bibitem{SedghiAnalyzingImpactVibrations2016}
Sedghi, B., Müller, M., and Dimmler, M., ``Analyzing the impact of vibrations
  on e-elt primary segmented mirror,''  991111 (08 2016).

\bibitem{leon-gilVibrationManagementGroundbased2025}
{Leon-Gil}, J., {Gonzalez-Cava}, J.~M., {Zamora-Jimenez}, A.,
  {Nu{\~n}ez-Cagigal}, M., and {Mendez-Perez}, J.~A., ``On the vibration
  management in ground-based telescopes. {{Use}} case for the {{European Solar
  Telescope}} ({{EST}}),'' {\em IFAC-PapersOnLine}~{\bf 59},  184--189 (Jan.
  2025).

\bibitem{conanWavefrontTemporalSpectra1995}
Conan, J.-M., Rousset, G., and Madec, P.-Y., ``Wave-front temporal spectra in
  high-resolution imaging through turbulence,'' {\em JOSA A, Vol. 12, Issue 7,
  pp. 1559-1570}  (July 1995).

\bibitem{barrettMaximumlikelihoodMethodsWavefront2007}
Barrett, H.~H., Dainty, C., and Lara, D., ``Maximum-likelihood methods in
  wavefront sensing: Stochastic models and likelihood functions,'' {\em JOSA
  A}~{\bf 24},  391--414 (Feb. 2007).

\bibitem{petitLEOSatelliteImaging2020}
Petit, C., Mugnier, L., Bonnefois, A., Conan, J.-M., Fusco, T., Levraud, N.,
  Meimon, S., Michau, V., Montri, J., Vedrenne, N., Velluet, M.-T., and
  F{\'e}tick, R., ``{{LEO}} satellite imaging with adaptive optics and
  marginalized blind deconvolution,'' in [{\em 21st {{AMOS Advanced Maui
  Optical}} and {{Space Surveillance Technologies
  Conference}}}{\nolinebreak\hspace{0.1em}]},  (Sept. 2020).

\bibitem{fetickIncludingPyramidOptical2023}
Fetick, R., Chambouleyron, V., and Taissir~Heritier, C., ``Including the
  pyramid optical gains into analytical models,''  9 pages, {Proceedings of
  AO4ELT7 conference 2023 - Editors: Thierry Fusco and Benoit Neichel} (2023).

\end{thebibliography}
\bibliographystyle{spiebib} 

\end{document}